\documentclass[fleqn,12 pt]{article}
\usepackage{amssymb,amsmath}
\usepackage{graphicx}
\usepackage{euscript}
\newcommand{\J}{\mathbb{J}}

\newcommand{\B}{\mathbb{S}}
\newcommand{\D}{\partial}

\begin{document}
\begin{titlepage}
\title{\textbf{Quantum-thermal self-diffusion as a
hydrodynamic mechanism for the fluctuation relaxation.}}
\author{\textbf{O.N. Golubjeva $^*$ and S.V. Sidorov}
\footnote {Peoples'  Friendship University of Russia,
\ 117198, Moscow, Russia. E-mail: ogol2013@gmail.com}}
\date {}
\maketitle

\begin{abstract}

We propose a generalization of quantum mechanical equations in the
hydrodynamic form by introducing, into the Lagrangian density, terms
taking into account the diffusion velocity at zero and finite
temperatures and the diffusion pressure energy of the warm vacuum.
Based on this, for the model of one-dimensional hydrodynamics, we
construct a system of equations that are analogous to the Euler
equations but with the inclusion of quantum and thermal effects.
They are a generalization of equations of the Nelson stochastic
mechanics. The numerical analysis of the behavior of solutions of
this system allows concluding that this system can be used to
describe the process of quantum-thermal fluctuation relaxation.


\bigskip
PACS\{02.60.Dc,\ 03.65.-w\}
\end{abstract}

\end{titlepage}

\section*{\small 1. Introduction}

Thermal fluctuations in hydrodynamics have been taken into account
during the fifty years; however, there is still no consistent
quantum statistical theory account for quantum and thermal effect
simultaneously \cite{For}. In this paper, we present a variant of
the approach to the construction of such a theory, starting from the
hydrodynamic form of quantum mechanics. For this purpose. we propose
to generalize it considering quantum-thermal diffusion, which
reflects the stochastic character of the environmental influence,
and the density of diffusion pressure energy, which exists at zero
and finite Kelvin temperatures. In this case, it is necessary to use
the generalization of the concept of thermal equilibrium to the case
of the simultaneous inclusion of stochastic influence of the quantum
and thermal types. As a result, for a one-dimensional model, we
obtain a system of hydrodynamic equations that is analogous to the
system of Euler equations, but differs from it, because it takes
quantum and thermal effects into account.

The hydrodynamic equations are traditionally derived either from the
statistical mechanics or from the kinetics, which use concrete
concepts of medium structure and interactions between its
components. Accordingly, hydrodynamic fluctuations are accounted for
by including, in the hydrodynamic equations, a random stress tensor
(together with the regular one), for which only one correlator is
given on the basis of the fluctuation-dissipation theorem (FDT).

At the same time, hydrodynamics is similar conceptually to
equilibrium thermodynamics, because it is also a modelless theory in
essence. Therefore, we propose to consider the theory of relaxation
of quantum-thermal fluctuations of the density and the drift
velocity at equilibrium with respect to the temperature as a
stochastic hydrodynamics. In this case, we can begin to derive the
corresponding equations by generalizing the hydrodynamic form of
quantum mechanics at zero temperature as a modelless theory to the
case of the explicit inclusion of self-diffusion in the cold and
warm vacuums. This allows, for the first time, extending the
hydrodynamic form of quantum mechanics to finite temperatures and
considering not only self-diffusion, but also the diffusion pressure
of the warm vacuum.

As a result, for the one-dimensional model, we obtain a system of
equations of stochastic hydrodynamics that is valid at any
temperatures. Its distinction consists in that it takes quantum
and thermal fluctuations into account nonadditively. Moreover,
we have managed to endow these equations with the form of the
equations of two-velocity hydrodynamics, which is a generalization
of the Nelson stochastic mechanics.

In our study, we rely on the results, which we obtained previously
in \cite{Sad2009}. In it, we developed the theory of
$(\hbar, k)$-dynamics, which allows introducing the consistent
quantum-thermal description of the thermal equilibrium state that
differs from the standard equilibrium thermodynamics and quantum
statistical mechanics (QSM).

The $(\hbar, k)$-dynamics is based on the idea of replacing the
classical thermostat model (as a set of classical oscillators)
with the distribution module $\theta_{cl} = k_BT$ with the
adequate quantum model (the quantum  thermostat,  or  the
"warm" vacuum,  which is the set of normal modes with all
frequencies $\omega$) with the distribution module
$\theta_{qu} = k_B\mathbb T$. Here, the quantity
\begin{equation}
\label{1}
\mathbb T
\equiv\frac{\hbar\omega}{2k_B}\coth\frac{\hbar\omega}{2k_BT}=
\varkappa\coth\frac{\varkappa \omega}{T}
\end{equation}
is called the effective temperature, the notation
$\varkappa=\hbar/2k_B$ is used for brevity.

The advantage of the characteristic $\mathbb T $ compared to the
Kelvin temperature $T$ is that it is never nonzero. This fact allows
considering the contact with the environment for $T\geqslant 0$ from
the general point of view, which is important in the case where
fluctuations of the quantum and thermal origins occurs
simultaneously and even in the case where only purely quantum
fluctuations exists. This quantity is accepted as a generalized
"mark" of the thermal equilibrium of an object being in contact with
the quantum thermostat (i.e. even at $T = 0$).

The main distinction of the $(\hbar, k)$-dynamics from QSM is that, under
the conditions of the equilibrium with the quantum thermostat, it describes
the object state not by the density matrix, but
by the complex wave function $\psi (q,\omega)$, whose amplitude and
phase are temperature-dependent. In the coordinate representation,
it has the form
\begin{equation}\label{ 2}
 \psi (q,\omega)=\left[2\pi(
 \Delta q )^2\right]^{-1/4}
 \exp\left \{-\frac{q^2}{4(\Delta q )^2}(1-i \alpha) \right\},
\end{equation}
where  $(\Delta q)^2$ is the coordinate variance and $\alpha$ is
the coefficient determining the phase.

At the same time, in the framework of the $ (\hbar,k) $-dynamics,
we have also introduced a new
macroparameter, namely, the effective influence of the quantum thermostat on the system as
the average of the quantum-thermal influence operator $\mathbb J= \overline{\hat j}$
\begin{equation}\label{3}
\mathbb J=\frac{\hbar}{2}\sqrt{\alpha^2+ 1}=\J^0 \sqrt{\alpha^2+ 1}.
\end{equation}
Here, $\mathbb J^0=\hbar/2$ is the limiting value of $\mathbb J$ at
the Kelvin temperature $T \rightarrow 0$ corresponding to the purely
quantum influence. In this case, the phase factor $\alpha$ vanishes,
which corresponds to the particular case of the real wave
function $\psi$. In the general case, the temperature dependence of
the effective influence is in the radicand of \eqref{3}.
If $\varkappa=\hbar/2k_B$ and the phase factor in formula \eqref{3}
$$
\alpha^2\equiv\sinh^{-2}\frac{\varkappa\omega}{T}
$$
are taken into account, the effective influence $\mathbb J$ \eqref{3} becomes
\begin{equation}\label{4}
\mathbb J=\frac{\hbar}{2}\coth\frac{\varkappa \omega}{T}=
\frac{\hbar}{2}\coth\frac{\hbar\omega}{2k_BT}.
\end{equation}

\section*{\small 2. Effective influence as a universal characteristic
of transport processes: Self-diffusion coefficient}

We first note that, in accordance with \cite {Sad2009}, the most important
thermodynamic parameters, namely, the effective temperature $\mathbb T$,
the effective internal energy $\mathbb U$, and the effective entropy
$\mathbb S$, are expressed in terms of the effective influence
$\mathbb J$ in the equilibrium case
\begin{equation}
\label{5}
\mathbb T=\frac {\omega}{k_B}\J,
\end{equation}
\begin{equation}\label{6}
\mathbb U=\omega \J, \;\;\;\;\mathbb S=-k_B\Bigl(1+\ln 2
\frac{\mathbb J}{\hbar}\Bigr).
\end{equation}

Introducing the limiting values of $\B$ and $\J$ in the forms
$\B^0=k_B$ and $\J^0=\dfrac\hbar 2$ as $T\rightarrow 0$,
which corresponds to the purely quantum influence, we can
establish the relation between  the effective influence and
the effective entropy
\begin{equation}
\label{7}
\mathbb J=\B^0\Bigl\{1+\ln\frac{\J}{\J^0}\Bigr\}.
\end{equation}
In connection with this, it would be natural to adopt the limiting
value of the ratio of two fundamental macroscopic quantities,
namely, the effective influence $\J$ and the effective entropy $\B$,
as a physical definition of the universal constant $\varkappa $:
\begin{equation}
\label{8}
\varkappa=\frac{\hbar}{2k_B}\equiv\lim_{T\to 0}
\frac{\J}{\B}= \frac{\J^0}{\B^0}.
\end{equation}

However, this does not exhaust the possibilities.
Effective transport coefficients (first of all, the
diffusion coefficient), which are typical of nonequilibrium
thermodynamics, can also be expressed in terms of this quantity.
Their stochastic nature is thus demonstrated. The latter is
obviously seen if the self-diffusion process is used as an example.
It occurs in the medium with the nonuniform density after the
equilibrium with respect to the temperature is established.

Indeed, it has already been shown in the theory of Brownian
motion at rather high temperatures \cite{Fuhr}
that, in this case
(for $t\gg\tau$), the uncertainty relation of the form
\begin{equation}
\label{9}
(\Delta p)\cdot(\Delta q) =  mD_T
\end{equation}
is valid. Here, $D_T$ is the coefficient of purely thermal
diffusion; in particular, \linebreak
$D_T = k_BT\tau/m$, where $\tau$ is the
relaxation time, for a free microparticle, while \linebreak
$D_T=k_BT/m\omega$
for a Brownian oscillator \cite{Sad2004}.

As shown in \cite{Sad2006}, the Schr\"odinger uncertainty relation
"momentum--coordinate" for a quantum oscillator
in the state of equilibrium with the thermal vacuum has the form
\begin{equation}\label{10}
(\Delta p)\cdot(\Delta q) = \J =
\frac{\hbar}{2}\coth\frac{\varkappa\omega}{T}.
\end{equation}
Comparing \eqref{9} and \eqref{10}, we rewrite this relation in the
form
\begin{equation}\label{11}
(\Delta p)\dot(\Delta q)=m\mathbb D.
\end{equation}
Then it is natural to call the quantity
\begin{equation}
\label{12}
\mathbb D= \frac{\hbar}{2m} \coth\frac{\varkappa\omega}{T}\equiv
\frac{\J}{m}
\end{equation}
the effective self-diffusion coefficient. We note that
Nelson \cite{Nel1967} called the quantity $\hbar/2m $ the quantum
diffusion coefficient $\hbar/2m=D_{qu} $, i.e., in the case of the
contact with the cold vacuum or, in other words, in the absence of

the thermal environmental influence.

It follows from \eqref{12} that the coefficient $\mathbb D$ acquires
the physical meaning of the effective influence per mass unit.
The respective limiting values of $\mathbb D$ at high and low
Kelvin temperatures are
\begin{equation}
\begin{split}
\label{13}
\mathbb D & \rightarrow D_T = \dfrac{k_BT}{m\omega}, \;\;\;
\mbox{high temperatures} \;\;\; k_BT\gg\hbar\omega/2,\\ 
\mathbb D & \rightarrow D_{qu}\equiv \dfrac{\hbar}{2m}, \;\;\;
\mbox{low temperatures}\;\;\;k_BT\ll\hbar\omega/2.
\end{split}
\end{equation}

Starting from relation \eqref{12}, one can
also introduce the other effective transport coefficients in terms
of $\J$, namely, coefficients of heat conductivity, shear viscosity,
and others that are important for nonequilibrium processes. Thus,
the majority of transport coefficients can be expressed in terms
of the effective self-diffusion coefficient $\mathbb D$, which can be,
in principle, measured in experiments.

As for the constant $\varkappa,$ it can be expressed in terms of
observed transport coefficients by means of relations of the
following type when analyzing particular experiments:
\begin{equation}
\label{14}
\varkappa=\left(\frac{\mathbb D}{\B/m}\right)_{min}=
\left(\frac{\eta_{ef}}{\B/V}\right)_{min}= ...,
\end{equation}
where $\B/m$ is the effective entropy of the mass unit, $\B/V$ is
the effective entropy of the volume unit, and $\eta_{ef}$ is the
effective shear-viscosity coefficient.

\section*{\small 3. Standard quantum mechanics in the hydrodymanic form}

In the nonrelativistic field form, the standard quantum mechanics
(at $T= 0$) can be obtained in the case where the action functional
variation becomes zero \cite{Feyn1967}:
\begin{equation}
\label{18-3}
\EuScript{S}=\int^{t_2}_{t_1}\;dt\int dq\;\mathcal
L_{\scriptscriptstyle{0}} [\psi^*;\psi].
\end{equation}
Here, $\mathcal L_{\scriptscriptstyle{0}} [\psi^*;\psi] $ is the
Lagrangian density for one spinless particle at $T=0$, and
$\psi(q,t) $ and $\psi^*(q,t)$ are the wave function and its
complex conjugate function; they have the meaning of independent
nonrelativistic fields. We restrict ourselves to the
one-dimensional case.

It is obvious that, in the general case, the functional
$\mathcal L_{\scriptscriptstyle{0}}
[\psi^*;\psi]$  must be chosen in the form
\begin{equation}\label{19-3}
\mathcal L_{\scriptscriptstyle{0}}[\psi^*;\psi]=\psi^*
(q,t)\left(i\hbar\frac{\partial}{\partial t}
+\frac{\hbar^2}{2m}\;\frac{\partial^2}{\partial q^2}\right)
\psi(q,t)-\psi^* (q,t)U(q)\psi (q,t),
\end{equation}
where the nonrelativistic limit of the Klein--Gordon operator is in
brackets on the right, and the potential energy operator$ U(q) $
characterizes the regular influence energy.

The independent variation of the action of form \eqref{19-3} with
respect to the field $\psi^*$ leads to the condition
\begin{equation}\label{20-3}
\int^{t_2}_{t_1}\;dt\int dq\frac{\delta\mathcal
L_{\scriptscriptstyle{0}}[\psi^*;\psi]} {\delta
\psi}=\int^{t_2}_{t_1}\;dt\int
dq\left(i\hbar\frac{\partial\psi}{\partial t}
+\frac{\hbar^2}{2m}\;\frac{\partial^2\psi}{\partial
q^2}-U(q)\psi\right)=0,
\end{equation}
which leads to the Schr\"odinger equation
\begin{equation}\label{21-3}
i\hbar\frac{\partial\psi}{\partial
t}=-\frac{\hbar^2}{2m}\;\frac{\partial^2 \psi}{\partial
q^2}+U(q)\psi.
\end{equation}

Accordingly, its complex conjugate equation is obtained when varying
the action of form \eqref{19-3} with respect to $\psi$ and differs from
formula \eqref{21-3}  by the replacement of $i$  with $-i$
and of $\psi$ with $\psi^*$.  We stress that the Schr\"odinger
equations for the complex wave functions $\psi$ and $\psi^*$ have t
he meaning of the Euler--Lagrange equations; in this case, wave
functions are always complex in the full-scale quantum mechanics.

We now represent the wave function in the form
\begin{equation}\label{22-3}
\psi(q,t)=\sqrt{\rho(q,t)}\exp\{i\theta(q,t)\},
\end{equation}
where $\rho (q,t)=|\psi(q,t)|^2.$ It could be possible to substitute
this expression, along with its complex conjugate expression,
directly in the Schr\"odinger equations for $\psi$ and $\psi^*$
and obtain the system of equations for the functions $\rho(q,t)$
and $\theta(q,t)$ that has long been known in the literature as
quantum mechanics in the hydrodynamic
form \cite{Feyn1967}, \cite{Bl1966}.

Because our aim is to construct modified hydrodynamics based on the
microdescription, we propose another approach to the problem. It
requires to develop the theory in the Lagrange formulation from the
beginning. Therefore, we start from transforming the Lagrangian
density $\mathcal L_{\scriptscriptstyle{0}}$  to variables that are
most suited to the hydrodynamic description. As functional arguments
of the Lagrangian density, we choose two independent real functions,
namely, the probability density $\rho$ and the phase $\theta$
instead of the complex wave functions $\psi$ and $\psi^*$. In
essence, they are close to the functions of the mass density $\rho_m$
and drift velocity $v\sim\dfrac{\partial \theta}{\partial q}$,
which are typical of standard hydrodynamics.

To do this, we replace the arguments in the Lagrangian density
\eqref{19-3} by substituting expression \eqref{22-3} and the
corresponding expression for $\psi^*$ in it. After the
substitution, we obtain
\begin{eqnarray}\label{23-3}
\mathcal L_{\scriptscriptstyle{0}}[\psi;\psi^*]=\mathcal
L_{\scriptscriptstyle{0}}[\rho;\theta]=-\hbar
\frac{\partial\theta}{\partial t}\rho\;-\frac{\hbar^2}{2m}
\left(\frac{\partial \theta}{\partial q}\right)^2\rho-
\frac{\hbar^2}{8m}\left(\frac{\partial \rho}{\partial q}
\right)^2\frac{1}{\rho}-\nonumber\\-U(q)\rho+i\frac{\hbar}{2}
\frac{\partial\rho}{\partial t}+\frac{\hbar^2}{2m}\cdot
\frac{\partial}{\partial q}\left( \frac 12
\frac{\partial \rho}{\partial q}+i\rho\;
\frac{\partial\theta}{\partial q}\right).
\end{eqnarray}
Here, the term containing $\dfrac{\partial\rho}{\partial t}$
cannot be taken into account, because it gives the zero contribution
when varying the action $\EuScript{S}$ of form \eqref{18-3} with  respect
to both $\theta$ and $\rho$ in what follows. The last term in
\eqref{23-3} is the total derivative with respect to $q$,
so that it can also be excluded from the definition of
$\mathcal L_{\scriptscriptstyle{0}}[\rho;\theta]$.
Therefore, as an expression for the Lagrangian density
$\mathcal L_{\scriptscriptstyle{0}}[\rho;\theta]$, we finally take
the following expression:
\begin{equation}\label{24-3}
\mathcal L_{\scriptscriptstyle{0}}[\rho;\theta]=-\hbar\;
\frac{\partial\theta}{\partial t}\;\rho\;
-\frac{\hbar^2}{2m}\left(\frac{\partial \theta}{\partial q}\right)^2
\rho\;-\;\frac{\hbar^2}{8m}\left(\frac{\partial \rho}{\partial q}
\right)^2\frac{1}{\rho}\;-U(q)\rho.
\end{equation}

Consistently varying the action $\EuScript{S}$ of form \eqref{18-3},
in which now $\mathcal L_{\scriptscriptstyle{0}}[\rho;\theta]$
has form \eqref{24-3}, with respect to the variable $\theta$
and $\rho$, we obtain the equations for the real functions
$\rho(q,t)$ and $\theta(q,t)$:
\begin{equation}
\label{25-3}
\frac{\partial\rho}{\partial t}+\frac{\partial}{\partial q}
\left(\rho\frac{\hbar}{m}\;\frac{\D\theta}{\D q}\right)=0,
\end{equation}
\begin{equation}\label{26-3}
\hbar\frac{\partial\theta}{\partial
t}+\frac{\hbar^2}{2m}\left(\frac{\partial \theta}{\partial q}
\right)^2+U(q)-\frac{\hbar^2}{8m}\left[\frac{1}{\rho^2}
\left(\frac{\partial\rho}{\partial q} \right)^2+2
\frac{\partial}{\partial q}\left(\frac{1}{\rho}\;
\frac{\partial \rho}{\partial q}\right)\right]=0.
\end{equation}
These equations coincide with the equations that could be obtained
for the functions $\rho$ and $\theta$ directly from the
Schr\"odinger equations. However, it is now clear that they have the
meaning of the Lagrange-Euler equations for the
action $\EuScript{S}$ of form \eqref{18-3} expressed in terms of the
variables $\rho$ and $\theta$.

It is assumed traditionally that Eq. \eqref{25-3} is the continuity
equation for $\rho (q,t)$. In turn, Eq. \eqref{26-3} is an analogue
of the Hamilton-Yacobi equation if the fact that the quantity
$\hbar\theta(q,t)$ has the dimensionality of action is taken into
account. In this case, the term in brackets in formula \eqref{26-3}
is sometimes treated as an additional energy of quantum
nature $U_{qu}(q)$ vanishing in the quasiclassical limit as
$\hbar\rightarrow 0$.

Of course, Eqs. \eqref{25-3}  and \eqref{26-3}
for $\rho$ and $\theta$ and the Schr\"odinger equations for
$\psi$ and $\psi^*$ are equivalent formally.  However, the
derivation of the quantum mechanical equations in hydrodynamic
form \eqref{25-3} and \eqref{26-3} directly from the principle of
least action is physically more preferable to construct stochastic
hydrodynamics. At the same time, to obtain the desired result, we
must solve the problem of the form of the Lagrangian density (which
remains unsolved), in which, in our opinion, the stochastic
influence of the environment (the quantum thermostat) must be
consistently taken into account.

\section*{\small 4. Quantum self-diffusion in the "cold" vacuum}

To reveal the possibility of
generalizing $\mathcal L_{\scriptscriptstyle{0}}[\rho;\theta]$,
we first consider the case of the cold vacuum. To do this, we
endow the second and third terms on the right in expression
\eqref{24-3} with the physical meaning. In accordance with
the terminology introduced by Kolmogorov \cite{Kol1933}
for Markovian processes in the general theory of stochastic processes
and used by Nelson \cite{Nel1967}
in its stochastic mechanics, we call the quantity
\begin{equation}\label{24}
v\equiv\frac{\hbar}{m}\;\frac{\D\theta}{\D q}
\end{equation}
the drift velocity. Accordingly, we call the quantity
\begin{equation}
\label{25}
u\equiv-D_{qu}\frac 1\rho\;\frac{\D\rho}{\D q}=
-\frac{\hbar}{2m}\;\frac{1}{\rho}\;\frac{\D\rho}{\D q}
\end{equation}
the diffusion velocity in the cold vacuum and stress its stochastic
quantum nature initially.

Using the velocities $v$ and $u$, we can write
formulas \eqref{24-3} -\eqref{26-3} in the form

$$\mathcal L_{\scriptscriptstyle{0}}[\rho,\theta]
=-\hbar\;\frac{\D\theta} {\D t}\;\rho-\frac m2(v^2+u^2)\rho-U\rho,
\eqno (21a)
$$

$$\frac{\D\rho}{\D t}\;+\frac{\D}{\D q}\;(\rho v)=0,\eqno (22a)$$
$$\hbar\frac{\D\theta}{\D t}+\frac m2\;v^2+U-\frac m2[u^2-
\frac{\hbar}{m}\frac{\D u}{\D q}]=0,\eqno (23a)$$
providing a possibility of generalizing the Lagrangian density.

It follows from formula (22a) that standard continuity equation (22)
is of quasiclassical character, because the probability flux density in it
depends only on the drift velocity  $v$, while the diffusion velocity $u$
generated by the stochastic influence of the cold vacuum is not taken into account
in it.

In connection with this, we recall that
the Fokker--Planck equation
\begin{equation}\label{26}
\frac{\D\rho}{\D t}+\frac{\D}{\D q}(\rho V)=0,
\end{equation}
which contains the total velocity of the probability flux density
\begin{equation}\label{27}
V=v+u,
\end{equation}
is the most general continuity equation in accordance with
Kolmogorov \cite{Kol1933}. We show that it allows describing the
approximation to the thermal equilibrium state because of
self-diffusion, including the case of the cold vacuum.

Attention is drawn to the fact that the combination $\dfrac{m}{2}(v^2+u^2)$
contained in expression (21à) for $\mathcal L_{\scriptscriptstyle{0}}
[\rho,\theta]$ is the sum
of independent contributions of the kinetic energies of
the drift and diffusion motions. At the same time, the probability flux
depends on the total velocity of form \eqref{27}.
In connection with this, to obtain the Fokker--Planck equation,
in expression (21à), the natural replacement of $(v^2+u^2)$ with $V^2$
must be performed, which allows taking into account the total
expression for the kinetic energy related to the probability flux.
Thus, even the standard quantum mechanics (at $T=0$) admits the
possibility to generalize.

Thus,  we generalize the Lagrangian density  $\mathcal
L_{\scriptscriptstyle{0}}[\rho,\theta]$ of form (21a) by means of
the corresponding replacement. Then we obtain
\begin{equation}
\label{28}
\tilde{\mathcal
L}_{\scriptscriptstyle{0}}[\rho,\theta]=-\hbar\;\frac{\D\theta}{\D
t}\;\rho-\frac m2V^2\rho-U\rho=\mathcal
{L}_{\scriptscriptstyle{0}}[\rho;\theta]-mvu\rho=\mathcal
{L}_{\scriptscriptstyle{0}}[\rho;\theta]+\frac{\hbar^2}{2m}\;
\frac{\D\theta}{\D q}\;\frac{\D\rho}{\D q}.
\end{equation}

Varying the action functional $\EuScript{S}$ of form \eqref{18-3}
with $\tilde{\mathcal L}_{\scriptscriptstyle{0}}[\rho,\theta]$
with respect to $\theta$ automatically leads to the Fokker--Planck
equation with the quantum diffusion coefficient $D_{qu}$
\begin{equation}\label{29}
\frac{\D\rho}{\D t}+\frac{\D}{\D q}(\rho V)=
\frac{\D\rho}{\D t}+\frac{\D}{\D q}\left(\rho\;\frac{\hbar}{m}\;
\frac{\D\theta}{\D q}\right)-D_{qu}\frac{\D^2\rho}{\D q^2}=0.
\end{equation}

At the same time, varying $\EuScript{S}$ with respect to $\rho$ barely
changes the Hamilton--Yacobi equation, in which the additional
insignificant term appears in comparison with (23a). As a result,
the analogue of Eq. (23a) becomes
\begin{equation}
\label{30}
\hbar\frac{\D\theta}{\D t}+\frac m2v^2+U-\frac
m2\left(u^2-\frac\hbar m \frac{\D u}{\D
q}\right)+\frac{\hbar}{2}\;\frac{\D v}{\D q}=0.
\end{equation}
Obtained equations \eqref{29} and \eqref{30} generalize Eqs.~(22a)
and (23a), which allows consistently taking the quantum stochastic
influence of the cold vacuum into account.

\section*{\small 5. Self-diffusion in the quantum thermostat
for $T\ne0$}

We now use the approach developed above to the description of
self-diffusion simultaneously taking quantum and thermal effect
into account. To do this, we introduce the temperature-dependent
Lagrangian density $\tilde{\mathcal L}_{\scriptscriptstyle{T}}[\rho,\theta]$
and require that it transform into the
expression $\tilde{\mathcal L}_{\scriptscriptstyle{0}}[\rho,\theta]$
of form \eqref{28} as $T\rightarrow 0$. To do this,
it suffices to replace the diffusion coefficient $D_{qu}$ with
$\mathbb D$ of form \eqref{8} in expression (28) for the diffusion
velocity and introduce the additional term $U_T(q)\rho$. The latter
takes into account the density of the diffusion pressure energy
because of the thermal stochastic environmental influence in the
expression for the Lagrangian density.

In our opinion, the expression for $U_T$ must have the form that is
analogous to the factor $-mu^2/2$ in the cold vacuum (21à). However,
it must be modified so that $U_T\rightarrow 0$ as $T\rightarrow 0$.
We introduce it as follows:
\begin{equation}
\label{31}
U_T(q)=-\frac m2\left[\frac{\alpha}{\Upsilon}\right]^2u_{ef}^2=-
\frac{\hbar^2}{8m}\;\alpha^{2}
\left(\frac{1}{\rho}\;\frac{\D\rho}{\D q}\right)^2,
\end{equation}
where the notations\quad
$\alpha^2\equiv\sinh^{-2}\dfrac{\varkappa\omega}{T}; \quad \Upsilon=
\coth\dfrac{\varkappa\omega}{T}$ and $u_{ef}\equiv -\mathbb D
\dfrac{1}{\rho}\dfrac{\D\rho}{\D q}$\quad are used. Here $u_{ef}$ is the
effective diffusion velocity in the warm vacuum. This quantity is
defined analogously to the velocity $u$ in \eqref{25}, but it is now
expressed in terms of the effective diffusion coefficient $\mathbb D
$ of form \eqref{12}.

Thus, as the Lagrangian density at $T\ne 0$, we choose the expression
\begin{equation}
\label{32}
\tilde{\mathcal
L}_T(\rho,\theta)=-\hbar\frac{\D\theta}{\D t}\;\rho -\frac
m2(v+u_{ef})^2\rho-U\rho-U_{\scriptscriptstyle T}\rho.
\end{equation}
For the convenience of the next variation, we rewrite
expression \eqref{32} in the explicit form in terms of the random
functions $\theta$ and $\rho$:
\begin{multline}
\label{33}
\tilde{\mathcal L}_T(\rho,\theta)=-\hbar\;\frac{\D\theta}{\D
t}\;\rho -\left\{\frac{\hbar^2}{2m}\;\left(\frac{\D\theta}{\D
q}\right)^2\rho-\frac{\hbar^2}{2m}\Upsilon\frac{\D\theta}{\D
q}\;\frac{\D\rho}{\D
q}+\frac{\hbar^2}{8m}\Upsilon^2\frac{1}{\rho}\;\left(\frac
{\D\rho}{\D q}\right)^2\right\}-\\-
U\rho-\frac{\hbar^2}{8m}\;\alpha^2
\frac{1}{\rho}\left(\frac{\D\rho}{\D q}\right)^2.
\end{multline}

Varying the action $\EuScript{S}$ with $\widetilde{\mathcal
L}_{\scriptscriptstyle{T}} $ of form \eqref{33} with respect to
$\theta$ leads again to the Fokker--Planck equation, which is
analogous to \eqref{29}, but with the replacement of $D_{qu} $ with
the effective diffusion coefficient $\mathbb D$ in it:
\begin{equation}\label{34}
\frac{\D\rho}{\D t}+\frac{\D}{\D q}\left(\rho\frac{\hbar}
{m}\frac{\D\theta}{\D q}\right)-\mathbb D\frac{\D^2\rho}{\D q^2}=0.
\end{equation}

Accordingly, varying $\EuScript{S}$ with respect to $\rho$ leads to
the Hamilton--Yacobi equation generalized to the case of the
stochastic influence of the thermal vacuum:
\begin{multline}
\label{35}
\hbar\frac{\D\theta}{\D t}+\frac{\hbar^2}
{2m}\left(\frac{\D\theta}{\D q}\right)^2+\frac{\hbar^2} {2m}
\Upsilon\;\frac{\D^2\theta}{\D
q^2}+U(q)-\\-\frac{\hbar^2}{8m}\Xi_T\left[\frac{1}{\rho^2}\;
\left(\frac{\D\rho}{\D q}\right)^2 +2\frac{\D}{\D
q}\left(\frac{1}{\rho}\frac{\D\rho}{\D q}\right)\right]=0,
\end{multline}
where the notation
\begin{equation}
\label{36}
\Xi_T=2 \Upsilon^2-1=2\coth^2\frac{\varkappa\omega}{T}-1,
\qquad \Xi_T\Big|_{T=0}\equiv\Xi_0=1
\end{equation}
was introduced for convenience.

In turn, obtained equations \eqref{34} and \eqref{35} generalize
Eqs. \eqref{29} and \eqref{30}, which allows consistently taking the
stochastic influence of the warm vacuum into account. It is
indirectly represented in the quantities $\mathbb D$, $\Xi_T$,
and $\Upsilon$, which are contained in these equations and are
dependent on the world constants $\hbar$ and $k_B$. This means
physically that both types of stochastic environmental influence are
taken into account simultaneously: the quantum one characterized by
the Planck constant $\hbar$ and the thermal one characterized by the
Boltzmann constant $k_B$.

Of course, the set of the Fokker--Planck equations \eqref{34} and
Hamilton--Yacobi equations \eqref{35} is a nontrivial generalization
of the Schr\"odinger equation. There are two ways of using them
later on. This system can be directly solved for unknown dissimilar
functions $\rho$ and $\theta.$ As we showed in \cite{Gol2011}
recently, this allows obtaining nonequilibrium wave functions whose
amplitudes and phases are temperature-dependent, and macroparameters
in nonequilibrium states can be calculated using them. But these
equations can also be modified by endowing them with the form of
equations of two-velocity stochastic hydrodynamics for the
characteristic velocities $v$ and $u$. As we will see, these
equations are a generalization of the corresponding equations of the
Nelson stochastic mechanics.

\section*{\small 6. One-dimensional model of two-velocity stochastic
hydrodynamics}

To modify the system of equations \eqref{32} and \eqref{33}, we can
make the next step and endow these equations with the form of
equations for variables of the same type, namely, the velocities $v$
and $u_{ef}$, which are typical of any Markovian processes. In this
case, we obtain the system of equations of two-velocity stochastic
hydrodynamics generalizing the equations of Nelson stochastic
mechanics to the case of the quantum-thermal environmental
influence.

We now show that Eqs. \eqref{34} and \eqref{35} really allow obtaining
equations of stochastic hydrodynamics in the most convenient form.
Taking the preceding into account, we are only dealing with the
one-dimensional model. To perform the corresponding transformation,
we first endow continuity equation \eqref{34}
with the form of the equation for the diffusion velocity, which,
in accordance with \eqref{25}, can be written in the form
$$u_{ef}=-\mathbb D \dfrac{\D\ln\rho}{\D q}.$$

To do this, we first transform Eq. \eqref{34}
by introducing $v$ and $u_{ef}$ explicitly in it and then multiply it
by $(-\mathbb D/\rho)$:
\begin{equation}\label{37 }
-\frac{\mathbb D }{\rho}\cdot\frac{\D\rho}{\D t}-\frac{\mathbb D
}{\rho}\left[\rho\frac{\D(v+u_{ef})}{\D q}+\frac{\D\rho}{\D
q}(v+u_{ef})\right]=0.
\end{equation}

We next differentiate the result with respect to $q$, change the
order of differentiation in the first term, and form (where it is
possible) $\log \rho,$ which allows introducing $u_{ef}$ everywhere.
As a result, we obtain
\begin{equation}\label{38 }
\frac{\D u_{ef}}{\D t}+\frac{\D}{\D q}(vu_{ef})+\frac{\D}{\D
q}u_{ef}^2- \mathbb D \frac{\D^2}{\D q^2}(v+u_{ef})=0.
\end{equation}

It can be endowed with a more elegant form
by forming the substantial derivative of the diffusion velocity $u_{ef}$,
which is typical of hydrodynamics:

\begin{equation}\label{39}
\frac{du_{ef}}{dt}\equiv\frac{\D u_{ef}}{\D t}+u_{ef}
\frac{\D u_{ef}}{\D q} =-\frac{\D}{\D q}(vu_{ef})-
\frac{\D}{\D q}\frac{u_{ef}^2}{2}+ \mathbb D_{ef}
\frac{\D^2}{\D q^2}(v+u_{ef}).
\end{equation}

To endow Eq. \eqref{35} with the explicit hydrodynamic form,
we also rewrite it in the variables $v$ and $u_{ef}$:
\begin{equation}\label{40}
\hbar\frac{\D\theta}{\D t}+\frac m2 v^2+\frac\hbar 2
\Upsilon\frac{\D v}{\D q}+ U(q)-\frac m2\Xi_T\cdot(u_{ef}^2-
\frac\hbar m\frac{\D u_{ef}}{\D q})=0.
\end{equation}

To exclude the function $\theta$, we differentiate Eq. \eqref{40} with
respect to $q$, change the order of differentiation in the first term,
and introduce the function $v$ explicitly in it in accordance with
\eqref{24}. As a result, we obtain
\begin{equation}
\label{41}
\frac{d v}{d t} \equiv \left(\frac{\D v}{\D t}+v\frac{\D v}{\D
q}\right)=- \frac 1m\frac{\D U}{\D q}+\Xi_T\frac{\D}{\D q}
\frac{u_{ef}^2}{2}-\frac{\hbar}{2m}\left(\Xi_T\frac{\D^2 u_{ef}}{\D
q^2}+\Upsilon\frac{\D v^2}{\D q^2}\right)
\end{equation}
by forming the substantial derivative
of the drift velocity $v$ in it too.

We recall that, as the velocity $v$, these equations contain only
the quantity $\Delta v$ generated by the stochastic influence.
In the case under consideration, as in the case $T=0$,
the expressions for $\rho$ and $\theta$ are related to the wave
functions of thermal correlated-coherent states, in which the
exponent of the exponential depends on $q^2$ \cite{Sad2009},
\cite{Gol2014}. It hence follows that the last terms in Eqs.
\eqref{39} and \eqref{41} containing the second  derivatives
of $v\equiv\Delta v$ and $u_{ef}$ with respect to $q$ vanish.

As a result, in the general case, the system of equations for
the one-dimensional model of two-velocity stochastic hydrodynamics
becomes
\begin{equation}\label{42}
\begin{cases}
\dfrac{d u_{ef}}{dt}\equiv \dfrac{\D u_{ef}}{\D t}+
\dfrac{\D}{\D q}\dfrac{u_{ef}^2}{2} =-\dfrac{\D}{\D q}(vu_{ef})-
\dfrac{\D}{\D q}\dfrac{u^2_{ef}}{2},\\
\dfrac{dv}{dt}=-\dfrac 1m \dfrac{\D U}{\D q}+\Xi_T
\dfrac{\D}{\D q} \dfrac{u^2_{ef}}{2}.\\
\end{cases}
\end{equation}
We note that, in the general case, proposed equations \eqref{42} are
valid for any temperature.

Both equations of this system take into account self-diffusion in
the warm vacuum, which is characterized by the coefficient $D_{ef}$
contained in $u_{ef}$. In addition, the right-hand side of the lower
equation of the system contains the gradient of diffusion pressure
energy density reflecting the stochastic quantum-thermostat
influence, including the case $T=0$, in addition to the gradient of
classical potential $U(q)$. The analogous contribution of the
diffusion pressure energy of the quantum thermostat is also
contained in the right-hand side of the upper equation
of \eqref{42}; in this case, it does not vanish even for $v=0$.

For the comparison with the equations of the Nelson stochastic mechanics
\begin{equation}
\label{43}
\left\{
\begin{array}{lcr}
\dfrac{\D u}{\D t}=-\dfrac{\D}{\D q}(vu),  \\
\dfrac{dv}{dt}=-\dfrac 1m  \dfrac{\D U}{\D q}+\dfrac{\D}{\D q}
\dfrac{u^2}{2}, \\
\end{array}
\right.
\end{equation}
which are valid only at $T=0,$ we consider system of
equations \eqref{42} in the case of the cold vacuum ($T=0$).

In this case, $u_{ef}$ transforms into $u$. In addition,
for the convenience of comparison,
in the upper equation of this system, we return to the
partial derivative with respect to the time; to do this,
we combine similar terms in formula \eqref{42}. As a result, we have
\begin{equation}
\label{44}
\begin{cases}
\dfrac{\D u}{\D t}=-\dfrac{\D}{\D q}(vu)-\dfrac{\D}{\D q} u^2,\\
\dfrac{dv}{dt}=-\dfrac 1m \dfrac{\D U}{\D q}+\dfrac{\D}{\D q}
\dfrac{u^2}{2}.
\end{cases}
\end{equation}

As is seen, the lower equations of systems \eqref{43} and \eqref{44}
are identical completely. However, these systems of equations differ
significantly, which is noticeable when comparing the upper
equations. As was expected, it is related to the fact that our
theory takes into account self-diffusion occurring even in the cold
vacuum. As a consequence of this, the equation for the diffusion
velocity contains the gradient of diffusion pressure energy density.

Below, we compare the solutions of system of equations \eqref{43}
and \eqref{44} to establish important physical distinctions
between them occurring at $T=0$. In fact, it is interesting
to know how the inclusion of the self-diffusion
in the cold vacuum ($T=0$) in Eqs. \eqref{44} affects the forms of the
obtained solutions compared with those of system \eqref{43}.

\section* {\small 7. Study of the solutions of Eqs. \eqref{43} and
\eqref{44}}

We note that, as shown in Sec. 6, system \eqref{42} was obtained by
taking self-diffusion at arbitrary temperatures into account, while
system \eqref{44} is valid only at $T=0.$ Therefore, only system
of equations \eqref{43}, which is a particular case of \eqref{42}
at $T=0,$ and system \eqref{44} can be compared correctly.

\subsection*  {\small 7.1. Apparatus for the determination
of the class of equations}

As is known,  the total derivative is given by
\begin{equation}
\label{01=45}
\frac{dv}{dt} = \frac{\partial v}{\partial q}\frac{\partial q}{\partial t}
+\frac{\partial v}{\partial t} = v\frac{\partial v}{\partial q} +
\frac{\partial v}{\partial t}.
\end{equation}
Now we rewrite Eqs. \eqref{43}  and \eqref{44}  completely in the
partial derivatives; for convenience, we
set $\dfrac{1}{m}\dfrac{\partial U}{\partial q} =
\widetilde{\alpha}(q)$. Then systems of equations \eqref{43} and
\eqref{44} become
\begin{equation}\label{02=46}
\begin{split}
\frac{\partial u}{\partial t} & + v\frac{\partial u}{\partial q} +
u\frac{\partial v}{\partial q} = 0,\\
\frac{\partial v}{\partial t} & + v\frac{\partial v}{\partial q} -
u\frac{\partial u}{\partial q} = \widetilde{\alpha}(q),
\end{split}
\end{equation}
respectively, and
\begin{equation}
\label{03=47}
\begin{split}
\frac{\partial u}{\partial t} & + (v+2u)\frac{\partial u}{\partial q}
+ u\frac{\partial v}{\partial q} = 0,\\
\frac{\partial v}{\partial t} & + v\frac{\partial v}{\partial q} -
u\frac{\partial u}{\partial q} = \widetilde{\alpha}(q),
\end{split}
\end{equation}
where $u$ is the diffusion velocity, $v$ is the drift velocity.
Equations \eqref{02=46}
and \eqref{03=47} are quasilinear systems of differential equations
of first order for two unknown functions $u(t,  q)$ and $v(t,  q)$
of two variables. We note that the main distinction of
system \eqref{02=46} from system \eqref{03=47} is that the
assumption of self-diffusion was used when deriving the latter.

In this   paper,   we   study   homogeneous   systems    \eqref{02=46}
\eqref{03=47}  and set $\widetilde{\alpha}(q)=0$.  The establishing of
the type of equation (elliptic,  hyperbolic,  or parabolic one) is the
most important fact determining the solution of these equations. As is
known,  to solve hyperbolic equations,  the concept of  characteristic
(integrals of a certain characteristic equation) is used. The elliptic
operator has no characteristics in the real domain,  and,  in general,
stationary  equilibrium  states  correspond  to  elliptic differential
equations in physics. Thus, the establishing of the class to which the
corresponding  system  belongs allows drawing the conclusion about the
character  of  solutions  of  the   equation   and   set   them   into
correspondence with a certain physical interpretation.

Starting from the foregoing, we begin our study.
Equations \eqref{02=46} and \eqref{03=47} are system of quasilinear
partial differential equations of first order for two unknown functions.
Therefore, following \cite{Courant}, we represent each of the equations
in systems \eqref{02=46} and \eqref{03=47} in the form
\begin{equation}\label{04=48}
\begin{split}
L_1=A_1u_t+B_1u_q+C_1v_t+D_1v_q,\\
L_2=A_2u_t+B_2u_q+C_2v_t+D_2v_q,
\end{split}
\end{equation}
where $A_i,\ B_i,\ C_i,\ D_i$ are known functions of the
variables $t,  q, u, v,\quad i = 1, 2$. We assume that all the
considered functions are continuous and have continuous derivatives
of required order. It is well known that the linear
combination $af_x + b f_y$ of partial derivatives of the function of
two variables $f(x,y)$  is the derivative in the direction specified
by the relations $\dfrac{dx}{dy}  = \dfrac{a}{b}$.  If $x(l),\ y(l)$
is a curve with $\dfrac{x_l}{y_l}  = \dfrac{a}{b}$, then $af_x +
bf_y$ is the derivative of the function $f$ along this curve. This
fact allows us to elegantly pass from systems \eqref{02=46}
and \eqref{03=47} of partial differential equations to the study of
algebraic equations.

We consider functions $u(t,q),\ v(t,q)$ for which the coefficients in
differential equations \eqref{04=48} depend only on $t$ and $q$.
We find the linear combination
\begin{equation}\label{003=49}
L = \lambda_1L_1 + \lambda_2L_2,
\end{equation}
such that the differential expression $L$ contains the derivatives
only along one direction. Such a direction that depends on the
point $(t, q)$ and on the functions $u(t,  q)$ and $v(t,  q)$ at
this point is called characteristic. Let this direction be specified
by the ratio $t_l:q_l$. As was mentioned above, the condition that
the function $u(t, q),\   v(t,   q)$ in the differential
expression $L$ are differentiated in this direction then looks as
follows
\begin{equation}\label{05=50}
\frac{\lambda_1A_1 + \lambda_2A_2}{\lambda_1B_1+\lambda_2B_2} =
\frac{\lambda_1C_1 + \lambda_2C_2}{\lambda_1D_1+\lambda_2D_2} =
\frac{t_l}{q_l},
\end{equation}
because the coefficients at the derivatives $u_t,\  u_q$ and $v_t,\
v_q$ in the expression $L$ are determined by the corresponding terms
in proportions \eqref{05=50}. Multiplying expression \eqref{003=49}
by $t_l$, we obtain
\begin{multline}
\label{06=51}
Lt_l=(\lambda_1A_1 +    \lambda_2A_2)u_tt_l    +    (\lambda_1B_1 +
\lambda_2B_2)u_qt_l + \\
(\lambda_1C_1     +      \lambda_2C_2)v_tt_l
+ (\lambda_1D_1 + \lambda_2D_2)v_qt_l =\\
(\lambda_1A_1 + \lambda_2A_2)(u_l  -  u_qq_l)   +   (\lambda_1B_1
+ \lambda_2B_2)u_qt_l + \\
(\lambda_1C_1 + \lambda_1C_2)(v_l-v_qq_l)  +
(\lambda_1D_1 + \lambda_2D_2)v_qt_l =\\
(\lambda_1A_1 + \lambda_2A_2)u_l + (\lambda_1C_1 + \lambda_2C_2)v_l -\\
[(\lambda_1A_1+\lambda_2A_2)q_l-(\lambda_1B_1+\lambda_2B_2)t_l]u_q -\\
[(\lambda_1C_1+\lambda_2C_2)q_l-(\lambda_1D_1+\lambda_2D_2)t_l]v_q =\\
(\lambda_1A_1 + \lambda_2A_2)u_l + (\lambda_1C_1 + \lambda_2C_2)v_l,
\end{multline}
because, in view of \eqref{05=50}, we have
\begin{multline}\label{07=52}
(\lambda_1A_1+\lambda_2A_2)q_l-(\lambda_1B_1+\lambda_2B_2)t_l =\\
(\lambda_1C_1+\lambda_2C_2)q_l-(\lambda_1D_1+\lambda_2D_2)t_l =0.
\end{multline}
Analogously, multiplying $L$ by $q_l$, we obtain
\begin{equation}
\label{06=53}
Lq_l=(\lambda_1B_1 + \lambda_2B_2)u_l + (\lambda_1D_1 +
\lambda_2D_2)v_l.
\end{equation}

If the functions $u(t,q),\ v(t,q)$ are solutions of system \eqref{03=47},
and the expression $L$ has the derivative in the direction $l$, which is
given by the ratio $t_l:q_l$, then, from \eqref{07=52}, it is easy to obtain
the system of two linear homogeneous algebraic equations for $\lambda_1$
and $\lambda_2$
\begin{equation}\label{08=54}
\begin{split}
\lambda_1(A_1q_l - B_1t_l) + \lambda_2(A_2q_l - B_2t_l) = 0,\\
\lambda_1(C_1q_l - D_1t_l) + \lambda_2(C_2q_l - D_2t_l) = 0.
\end{split}
\end{equation}

System \eqref{08=54} has a nontrivial solution if it has the determinant
that is equal to zero, i.e.,
\begin{equation}
\label{55}
\begin{vmatrix}
A_1q_l - B_1t_l & A_2q_l - B_2t_l\\
C_1q_l - D_1t_l & C_2q_l - D_2t_l
\end{vmatrix} = 0,
\end{equation}
which is convenient to write in the quadratic form
\begin{equation}
\label{09=56}
at_l^2 - 2bt_lq_l + cq_l^2 = 0,
\end{equation}
where $a = [BD],\ 2b = [AD] + [BC],\ c = [AC],\ [XY] =
X_1Y_2-X_2Y_1$.

Depending on the sign of the determinant of form \eqref{09=56},
it is possible to classify the equations as follows.
\\ 1. If $b^2 - ac < 0$, then quadratic form \eqref{09=56} is nonzero
for any real $t_l,\ q_l$; and, consequently, there is no real
characteristic direction, and the system of differential equations
belongs to the elliptic type.\\
2. If $b^2 - ac  >  0$, then two characteristic
directions specified by the ratio $t_l : q_l$ exist at each point;
they correspond to two different roots $\lambda_1$ and $\lambda_2$
of the quadratic form \eqref{09=56}. In this case, the system
of differential equations belongs to the hyperbolic type.
\\
3. In the case where $b^2-ac  =  0$, expression \eqref{09=56} has
one root of multiplicity 2,  and there is one degenerate direction
corresponding to this root; the system of differential equations
belongs to the parabolic type.

\subsection*{\small 7.2. Study of Eqs. \eqref{02=46} and \eqref{03=47}}

Starting from the foregoing, we analyze
the class of system of Nelson equations \eqref{02=46}. For this system,
the coefficients have the form
\begin{equation}
\label{10=57}
\begin{split}
A_1 & =1,\ B_1 = v,\ C_1=0,\ D_1=u,\\
A_2 & =0,\ B_2 = -u,\ C_2=1,\ D_2=v.
\end{split}
\end{equation}
It hence follows that
\[b^2 - ac = - u^2 < 0.\]
Thus, the hydrodynamic system of Nelson equations is elliptic and
cannot be used to study the fluctuation evolution in the
quantum-mechanical description of the system.

For system \eqref{03=47}, the coefficients have another form
\begin{equation}\label{10=58}
\begin{split}
A_1 & = 1,\ B_1 = 2u+v,\ C_1=0,\ D_1=u,\\
A_2 & = 0,\ B_2 = -u,\ C_2=1,\ D_2=v,
\end{split}
\end{equation}
which leads to
$$
b^2 - ac = 0.
$$
The latter demonstrates that system of equations \eqref{03=47}
is parabolic, i.e., the system of the evolution type and,
consequently, can be used to describe the evolution of
perturbations appearing in the case of fluctuations.

We find the characteristic direction for system \eqref{03=47}.
In accordance with the foregoing, we form the linear combination of
two equations of this system and require that it contain
the derivatives of the functions $u(q,t)$ and $v(q,t)$ only
in one direction $u_l,\  v_l$ that is given by $(t_l, q_l)$:
\begin{equation}
\label{301=59}
L = u_t+(2u+v-\lambda u)u_q + \lambda v_t + (u+\lambda v)v_q = 0.
\end{equation}

Then, in accordance with \eqref{05=50}, the conditions determining
the direction $(t_l, q_l)$ looks as follows:
\begin{equation}
\label{302=60}
\begin{split}
q_l = & (2u + v -\lambda u)t_l,\\
\lambda q_l = & (u+\lambda v)t_l,
\end{split}
\end{equation}
whence
\begin{equation}\label{303=61}
(\lambda - 1)^2 = 0.
\end{equation}

Consequently, in problem \eqref{03=47}, there is one characteristic
direction determined by the condition
\begin{equation}
q_l = (u + v)t_l.
\label{304=62}
\end{equation}
From \eqref{301=59} -- \eqref{304=62}, we obtain the characteristic
equation for $u$ and $v$
\begin{equation}
u_l + v_l = 0.
\label{305=63}
\end{equation}

The meaning of Eq. \eqref{304=62} is that the characteristic in the
$(q,  t)$ plane represents the motion of possible perturbations
whose velocity \begin{equation}\label{306=64}
\frac{dq}{dt} = u + v
\end{equation}
is the sum of the drift and diffusion velocities.

In the context of the use of this system, we keep in mind the
following fact. The fluctuations of parameters, for example,
of the temperature, the density, and the pressure, necessarily
produce perturbations of the variables $u$ and $v$ in the
hydrodynamic equations. The evolution of these perturbations can
be described and studied using system of equations \eqref{03=47}.\\

\subsection*{\small 7.3. Numerical simulation
of the solutions of Eqs. \eqref{02=46} and \eqref{03=47}.}

Our above considerations are illustrated using the numerical
simulation of the solution of systems \eqref{02=46} and
\eqref{03=47}. The model equation for these systems
is the transport equation written in the vector form:
\begin{equation}
\label{11=65}
\frac{\partial y}{\partial t}+A(y)\frac{\partial y}{\partial q} = f,
\end{equation}
where $y  =  (u,  v)^{\text  T},\  A \in\Bbb R^{2\ltimes 2}$ is the
system matrix, and $f\in\Bbb R^2$. For systems \eqref{02=46}
and \eqref{03=47}, we consider the Cauchy problem
\begin{equation*}
\begin{split}
y(q, 0) = y_0(q),\\
q\in\Bbb R,\quad t > 0,
\end{split}
\end{equation*}
on the real axis.  To solve the problem in the $(q, t)$ plane, we
use the mesh
\begin{equation*}
\begin{split}
\omega_{h\tau} = & \omega_h \times \omega_{\tau},\\
\omega_h = & \{q_k = kh,\ k = 0, \pm 1, \pm 2, ...\},\\
\omega_{\tau} = & \{t_n= n\tau,\  n=0, 1, 2, ...\},
\end{split}
\end{equation*}
with the step $h$ with respect to $q$  and with the step $\tau$ with
respect to $t$. The solution of the problem was studied using the
implicit three-layer scheme
\begin{equation}
\label{12=66}
\frac{3y_k^{n+1}-4y_k^n + y_k^{n-1}}{2\tau} +
A(y)\frac{y_{k+1}^{n+1}-y_{k-1}^{n+1}}{2h}  = \varphi,
\end{equation}
with the approximation order $O(\tau^2+h^2)$. Problem \eqref{12=66} was
solved by the iteration method. We show that scheme \eqref{12=66} is
absolutely stable. For the homogeneous scalar equation
\begin{equation*}
\frac{\partial y}{\partial t}+ a(y)\frac{\partial y}{\partial q} = 0,
\end{equation*}
we seek a solution of problem \eqref{12=66} in the form
\begin{equation}\label{13=67}
y_k^n = \eta^n e^{ikh\theta},
\end{equation}
where $i=\sqrt{-1},\ \theta\in\Bbb R$. Substituting \eqref{13=67} in
the equation
\begin{equation*}
\frac{3y_k^{n+1}-4y_k^n + y_k^{n-1}}{2\tau} +
a(y)\frac{y_{k+1}^{n+1}-y_{k-1}^{n+1}}{2h} = 0,
\end{equation*}
after simple transformations, we obtain the equation for $\eta$
\begin{equation}
\label{14=68}
\mu\eta^2-4\eta + 1 = 0,
\end{equation}
where $\mu = 3 + 2a\gamma i\sin\theta,\ \gamma=\dfrac{\tau}{h}$.

We find the ensemble of points $G$ of the complex plane $\mu = r+is$
for which the absolute values of the roots of Eq. \eqref{14=68}
do not exceed unity. The boundary of the domain $G$ is the set of
points $\mu$ for which $\vert\eta\vert < 1$. We express the parameter
$\mu$ in Eq. \eqref{14=68} in terms of the variable $\eta$
\begin{equation*}
\mu = \frac{4}{\eta} - \frac{1}{\eta^2}.
\end{equation*}
It is obvious that if $\vert\eta\vert$ = 1, then, setting $\eta =
e^{-i\varphi}$, we obtain
\begin{equation*}
\mu = 4e^{i\varphi} - e^{2i\varphi}.
\end{equation*}
If the argument $\varphi$ is varied from $0$ to $2\pi$, then the
points $\mu$ describe the closed curve $\Gamma$; it is convenient to
represent the equation of this curve in the plane $\mu = r + is$ in
the parametric form
\begin{equation}
\label{15=69}
\begin{split}
r = 4\cos\varphi -\cos 2\varphi,\\
s = 4\sin\varphi -\sin 2\varphi.
\end{split}
\end{equation}
It can be seen from \eqref{15=69} that the curve $\Gamma$ is
symmetric with respect to the real axis $r$. It intersects the
axis $r$ at the points $\mu(0) = 3$ and $\mu(\pi) = -5$. At these
points, the derivatives
\begin{equation*}
\frac{ds}{dr} = \frac{2\cos\varphi - \cos 2\varphi}{\sin 2\varphi -
2\sin\varphi}, \qquad \frac{d^2s}{dr^2} =
\frac{-3}{4\sin\varphi (1- cos\varphi)^2}
\end{equation*}
are not determined. In this case, the second derivative of the
curve $\Gamma$ is negative for $0<\varphi  <\pi$ and positive
for $\pi < \varphi <2\pi$.  This is evidence of the fact that the
closed curve $\Gamma$ is convex upward in the upper
half-plane $\mu$  and convex downward in the lower half-plane.
Consequently, the domain inside the closed curve $\Gamma$ is convex
(Fig.~\ref{no1}).  In this case, the straight line $\mu=3+is$ touches
the curve $\Gamma$ at the point $\mu=3$. The other points of this
straight line lie in the domain located outside this curve $\Gamma$.
We show that the condition $\vert\eta\vert <1$ is satisfied in this
domain, and the set of points lying outside the curve $\Gamma$ is
the stability domain of scheme \eqref{12=66}.
\begin{figure}[h]
\center\includegraphics[bb= 0 0 460 330]{ris1.bmp}
\caption{The border of stability of cheme \eqref{12=66}.}
\label{no1}
\end{figure}

Indeed, we consider the solution of the equation
\begin{equation*}
(3+is)\eta^2-4\eta + 1 = 0
\end{equation*}
for $0< s< 1$. Then one of the roots
\begin{equation*}
\eta = \frac{2+\sqrt{1-is}}{3+is}
\end{equation*}
corresponding to the maximum value of the modulus $\vert\eta\vert$
can be written in the form
\begin{equation*}
\eta = \frac{9-\dfrac{s^2}{2}-\dfrac{9is}{2}}{9+s^2}+O(s).
\end{equation*}
The modulus of this value is
\begin{equation*}
\vert\eta\vert = \frac{\sqrt{81+\dfrac{45s^2+s^4}{4}}}{9+s^2}=
\sqrt{1-\frac{27s^2+s^4}{4(9+s^2)}} = 1-O(s^2)<1.
\end{equation*}

Thus, the set of points $\mu=3+2a\gamma i\sin\theta$ lies completely
in the stability domain, where $\vert\eta\vert \leq   1$.
Consequently, scheme \eqref{12=66} is absolutely stable and is
independent of the quantity $\gamma= \dfrac{\tau}{h}$ as it shown in
Fig.~\ref{no1}.
In the numerical simulation, we studied the solution for a certain
perturbation given in the initial condition in a neighborhood of the
point $q = 0$, which is one of the variables of the initial
homogeneous condition. This perturbation in the initial condition
models the result of the fluctuation of the physical-system
macroparameter. The calculations carried out numerically using the
proposed implicit scheme showed the following result. Any (even
arbitrarily small) perturbation in the system of Nelson equations
leads to an unbound growth of the variables $u$ and $v$. This is a
consequence of the fact that the Nelson equations belong to the
elliptic type and can be used to describe stationary processes such
as flow over bodies by a flux, problems of electrostatics,
stationary problems of gravitation theory.

When solving system of equations \eqref{44}, we obtained the
solution of the form of running perturbation wave with respect to
the spatial coordinate. In this case, the evolution of the
perturbation itself is observed along with the perturbation
displacement. Figure shows the numerical solutions of
system \eqref{44} for different instants of time:\quad  a) $t = 0$,
\quad b) $t = \tau$,\quad c) $t = 20\tau$,\quad d) $t = 50\tau$,
where  $\tau$ is the integration step with respect to the time.

\subsection* {\small 7.4. Discussion of the results of numerical
simulation}

When stating  the  problem of numerical simulation,  we start from the
fact the  fluctuations  of  macroparameters  of  the  physical  system
(temperature, density, and pressure) must produce perturbations of the
drift $v$  and  the  diffusion  $u$  velocities  in  the  hydrodynamic
equations.   Therefore,  we  intended  to  comparatively  analyze  the
behavior of the small perturbation of the quantities $v$  and  $u$  as
the  final aim of the numerical study of system of Nelson hydrodynamic
equations \eqref{43} and system \eqref{44} obtained in this paper.  In
other  words,  our problem was to clarify to what extent each of these
systems of equations is appropriate for  describing  the  dynamics  of
quantum-thermal   fluctuations.   To  answer  this  question,  a  weak
perturbation with respect to the variable $u$ in a neighborhood of the
point  $q_0$  was  introduced  as  an  initial condition in systems of
equations \eqref{02=46} and \eqref{03=47},  which are most  convenient
for  the  analysis.  The  numerical calculations carried out using the
above-mentioned implicit scheme showed the following.

1. In the Nelson system of equations, any, even an arbitrarily
small perturbation leads to an unbound increase in the variables $u$
and $v$. This is a consequence of the fact that the Nelson equations
are of the elliptic type and are intended to describe stationary
processes such as a flow around bodies, electrostatic problems,
stationary problems of gravitation theory, etc.

2. On the contrary, system of equations \eqref{03=47} demonstrated
a quite different result, namely, the fluctuation evolutions
in time and space, as shown in Fig.~\ref{no2}.
\begin{figure}[h]
\center\includegraphics[bb= 0 0 450 230]{ris2.bmp}
\caption{Results of numerical solution of system \eqref{03=47}.}
\label{no2}
\end{figure}

The initial conditions of the solution of problem \eqref{11=65} at the
instant  of  time t = 0 are given in Fig.~\ref{no2}a.  The behavior of
the variables $u$ and $v$ at the  subsequent  instants  are  shown  in
Figs.~\ref{no2}b--~\ref{no2}d.  It  follows from them that the initial
perturbation with respect to u decreases very rapidly.  Its  amplitude
decreases  significantly  even  during  the  period  $t = \tau$,  and,
moreover,  it does not remain localized at  the  point  $q_0$  and  is
dispersed   over   the   space,   moving   along  the  coordinate  $q$
(Fig.~\ref{no2}b).  In this case,  the response to the perturbation of
the  diffusion  velocity  $u$ also affects the drift velocity $v$.  We
note that its graph is two-hump. This reflects the fact that diffusion
occurs along the flow and in the opposite direction.

Figures ~\ref{no2}c  and  ~\ref{no2}d  shows  the  solutions of system
\eqref{11=65};  they demonstrate the  perturbation  evolution  at  the
subsequent   instants   of  time  $t  =  20\tau$  and  $t  =  50\tau$,
respectively.  This evolution is manifested by the further decrease in
the  amplitudes of the perturbation with respect to both variables $u$
and $v$,  in the spread of perturbations over  the  space  with  their
simultaneous  movement  in  the  space,  which  is caused by the drift
velocity.  It can be interpreted as a continuation of the decrease  in
the  amplitude  and  the  further  motion of the weakened perturbation
caused by the drift velocity.

\section*{\small 8.  Conclusion}

Fenyes \cite{Fen1952} was, probably, the first to put forward the idea
of using the Lagrangian density $\mathcal{L}[\rho;\theta]$ in  quantum
theory.  The  Fokker-Planck  equation containing the total probability
flux velocity $V$ with the diffusion coefficient  either  $D_{qu}$  or
$D_T$  followed from the expression proposed by him.  However,  he did
not introduce the generalized diffusion coefficient $D_{ef}$.  We note
that  he  obtained  another  equation of motion of the Hamilton-Yacobi
type in this paper. Moreover, he there stated that the obtained system
of  equations  for  the  functions  $\rho$  and $\theta$ is completely
equivalent to the Schr\"odinger equations for the functions $\psi$ and
$\psi^*$,  in  spite  of  the  fact  that,  unlike  the  Schr\"odinger
equation, the Fokker-Planck equation leads to the irreversibility.

Unlike \cite{Fen1952}, in our approach, we have consistently taken
into account quantum-thermal fluctuations and the density of
diffusion pressure energy related to the stochastic influence
of the environment (the quantum thermostat) for $T\geqslant0$.
As a final result, we endowed the Fokker-Planck and Hamilton-Yacobi
equations with the form of system of equations \eqref{42}
for the one-dimensional model of two-velocity stochastic hydrodynamics.
The self-diffusion coefficient in them is determined by the
effective environmental influence that is dependent on the
funcdamental constant $\varkappa = \dfrac{\hbar}{2k_B}$.

In our  opinion,  on  this  way,  in  the  future,  it  is possible to
construct the full-scale stochastic hydrodynamics taking into  account
not  only  self-diffusion  but also shear viscosity and then use it to
describe interesting media such as nearly perfect fluids (NPF).  To do
this,  it  is  necessary to pass from the lower equation in \eqref{42}
for the drift velocity to the equation that is a generalization of the
Navier-Stokes  equation  to  the case in which self-diffusion is taken
into account.

It follows from our analysis that the self-diffusion coefficient
$D_{ef}=\J/m$ is, probably, the most adequate characteristic of
transport phenomena and is important for describing dissipative
processes in NPF. To-day, it is possible to determine it
experimentally by studying the diffusion of massive quarks in the
quark-gluon plasma obtained during the collision between heavy ions.

The numerical analysis of the behavior of the solutions
of the particular case of system \eqref{42} in form \eqref{44}, which
is valid at $T=0$, showed that these equations illustrate the
perturbation relaxation. Thus, self-diffusion can be regarded
as a hydrodynamic mechanism for the relaxation of quantum-thermal
fluctuations. We intend to study the behavior of the solution of
system \eqref{42} in the general case later on.

Thus, we suggest that the hydrodynamic approach to quantum theory
proposed in this paper allows, in principle, studying quantum-thermal
fluctuations by means of the obtained hydrodynamic equations.

We are grateful to N.F.Shul'ga, I.M. Mrygloda, A.G. Zagorodnii, I.V.
Volovich, N.M. Plakida, Yu.P. Rybakov, and also the participants of
scientific seminars guided by them in Kiev, Kharkov, Samara, and
Moscow for the fruitful discussion of the results presented above.

We must say out of the sence of duty that the initial plan,
the problem statement, and the key ideas were proposed by our teacher
and co-author A.D. Sukhanov to the blessed memory of whom this
paper is dedicated.

\end{document}